# Room temperature manipulation of long lifetime spins in metallic-like carbon nanospheres

Bálint Náfrádi[1*], Mohammad Choucair[2*], Klaus-Peter Dinse[3], László Forró[1]

[1]Institute of Condensed Matter Physics, École Polytechnique Fédérale de Lausanne, Lausanne 1015, Switzerland.

[2]School of Chemistry, University of Sydney, Sydney 2006, Australia.

[3]Institut für Experimentalphysik, Freie Universität Berlin, Arnimallee 14, 14195 Berlin, Germany.

[*]Correspondence to: balint.nafradi@epfl.ch, mohammad.choucair@sydney.edu.au.

**The time-window for processing electron spin information (spintronics) in solid-state quantum electronic devices is determined by the spin-lattice ($T_1$) and spin-spin ($T_2$) relaxation times of electrons[1-3]. Minimising the effects of spin-orbit coupling and the local magnetic contributions of neighbouring atoms on $T_1$ and $T_2$ at room temperature remain substantial challenges to practical spintronics[4-7]. Here, we report a record-high conduction electron $T_1=T_2$ of 175 ns at 300 K in 37 nm ± 7 nm carbon spheres, which exceeds by far the highest values observed for any conducting solid state material of comparable size. The long $T_1=T_2$ is due to quantum confinement effects, to the intrinsically weak spin-orbit coupling of carbon, and to the protecting nature of the outer shells of the inner spins from the influences of environmental disturbances. Following the observation of spin polarization by electron spin resonance, we controlled the quantum state of the electron spin by applying short bursts of an oscillating magnetic field and observed coherent oscillations of the spin state. These results demonstrate the feasibility of operating electron spins in conducting carbon nanospheres as quantum bits at room temperature.**

Electron spin states are an attractive realization of a quantum bit (qubit) as they can undergo a transition between the spin-up and spin-down quantum states[8]. The most commonly used technique for manipulating electron spin is electron spin resonance (ESR)[9]. ESR is the physical process whereby electron spins are polarized in an external magnetic field $\mathbf{B_0}$ and rotated by an oscillating magnetic field $\mathbf{B_1}$ (perpendicularly to $\mathbf{B_0}$, of frequency $f$) which is resonant with the spin precession frequency in an external magnetic $f = g\mu_B\mathbf{B_0}/h$ ($\mu_B$ is the Bohr magneton and *g* the electron spin *g*-factor ). ESR is the result of the coherence of the precession of electrons over the spin-relaxation times $T_1$ and $T_2$[10]. $T_1$ is characterised by a number of spin-lattice relaxation process that depend on the spin-orbit coupling to connect the spin of an electron with the lattice vibrational spectrum of the solid. The second important relaxation time $T_2$ is set by the probability of spin-spin relaxation. $T_2$ is concerned with the local magnetic field contribution by one magnetic atom on others and represents the phase coherence of a set of spins. The dominant relaxation time is the shorter of $T_1$ and $T_2$. In magnetically homogenous itinerant systems (*e.g.* metals), the condition $T_1=T_2$ is often met and represents the longest period of time that in-phase precessing electron spins and magnetisation can propagate as a uniform mode[11].

Electron spin states therefore need to be robust against decoherence. The feasibility of applications involving classical or quantum information processing, is hence critically dependent on $T_1$ and $T_2$ relaxation times[12]. The prerequisite for $T_1$ and $T_2$ relaxation times is ~100 ns, as this is the state of the art lower-bound for signal processing times in quantum electronic devices[13,14].

Advances in the fields of inorganic[15,16] and molecular[17,18] quantum dots have made the electron spin system promising for practical application in spintronic and quantum information processing. Current research activities on spin based qubits are divided into two distinct research directions with limited overlap and are based on the materials employed: inorganic materials, usually semiconducting structures forming quantum dots, and carbon-based molecules. The qubits utilised in these materials are also different: inorganic materials containing localised spins of doped ions or conduction electron spin confined on quantum dots, whereas organic materials utilize localised paramagnetic spins confined on molecules.

Due to the radical differences in the materials and the electron systems employed for spin qubit manipulation, the major challenges to realising devices are also distinct. For solid-state inorganic materials the strong spin-orbit interaction enforced by heavy nuclei in materials like semiconductors[5,6,19] and metals[7,20,21] strongly requires the pursuit of structural perfection and spin manipulation at low temperatures (40 mK to 100 K). This has resulted in well-defined materials that currently set the benchmark for long $T_1$ relaxation times exceeding seconds[4]. Unfortunately, at room temperatures a number of intrinsic spin-lattice relaxation processes and phonon modes induce spin decoherence and diminish $T_1$ and $T_2$.[22]

Consequently, for metal nanoparticles[21] $T_1$ and $T_2$ shortens to 10–40 ps and for bulk semiconductors[23,24] to 1–4 ns.

Molecular qubits can be produced on an industrial scale and are readily processed.[18] These spin-bearing organic molecules often have low spin orbit coupling, possess many non-degenerate spin transitions, and can be chemically modified. These attributes have led to the potential of molecular compounds for both spintronics and quantum information processing. However, the dipole-dipole interactions between molecules and the hyperfine interaction of the molecule requires diluting the spin containing molecules and isotope engineering of the host system to reduce magnetic inhomogeneity. It also needs low temperatures (<100 K) for spin manipulation to dampen molecular vibrations which cause spin decoherence.

Attempts to combine the advantages of inorganic and molecular qubit materials have led to electronically insulating systems like nitrogen-vacancy (N-V) center nanodiamonds[25] and N@$C_{60}$.[14] Long microsecond $T_1$ and $T_2$ of localised electrons in these materials can be achieved and manipulated at ambient conditions. Magnetic inhomogeneity in these materials is accounted for by isotopic engineering constituent $^{13}C$ and $^{14,15}N$ nuclei, however, the localised electron spins themselves generate fluctuating magnetic fields, posing limitations on the qubit density. This limitation is specific to localised electron systems and it is absent for conduction electrons where the motional narrowing process negates this effect[26].

Carbon nanotube based quantum dots have been prepared with an electron spin dephasing time of ~3 ns at millikelvin temperature[27]. The shortening of $T_2$ in these carbon nanotubes is attributed to an increase in the spin orbit coupling induced by the high curvature of the carbon sheets[28]. Graphene on the other hand is comprised entirely from a single-atom thick crystalline layer of carbon, is flat (or at best not curved), and conducting[29,30]. Experiments on single- and few-layer graphene have demonstrated $T_1 = T_2$ of 1–4 ns at room temperature[31]. However, this value is still far below expectations for practical use and possibly limited by physisorption and chemisorption of molecular oxygen or other species onto the unprotected carbon surface[32]. Non-carbon based materials have also been developed, including heterometallic molecular nanomagnets comprising of isotopically engineered and diamagnetically diluted Cr$_7$M (M = Ni, Mn, Zn) that demonstrate the prerequisite for quantum processing applications at low temperature (5 K)[33].

These possible solutions demonstrate that there are clear and established trade-offs to be considered regarding the feasibility of applying a qubit material system: a conducting material of light atomic weight constituents that meets the prerequisite $T_1$ and $T_2$ at room temperature would permit real spintronics and quantum processing applications. Such a material would combine the best aspects of both inorganic and molecular spin qubit schemes.

Here, we report the observation of $T_1=T_2=175$ ns at room temperature (300 K) in conducting metallic-like carbon nanospheres via ESR. Furthermore, we could coherently controlled the quantum state of the electron spin confined on the carbon nanospheres at 300 K. This remarkable result we can put in context by comparing it to other promising systems which could carry the quantum bit: N@C60, N-V centre in nanodiamond or P doped Si (Fig. 1). In the carbon nanospheres the electron spins are highly delocalised and polarizable (Pauli-paramagnetic) over the volume of the nanosphere, rather than highly confined to a small path on the order of the electron orbit, as would be the case for a localised electron system. This has the advantages that there is no need for isotopic engineering of the carbon nuclei to supress magnetic inhomogeneity, and an absence of fluctuations in local magnetic fields of localised spins. The material thus allows a higher density packing of qubits to be, in principle, achieved over other promising qubits.

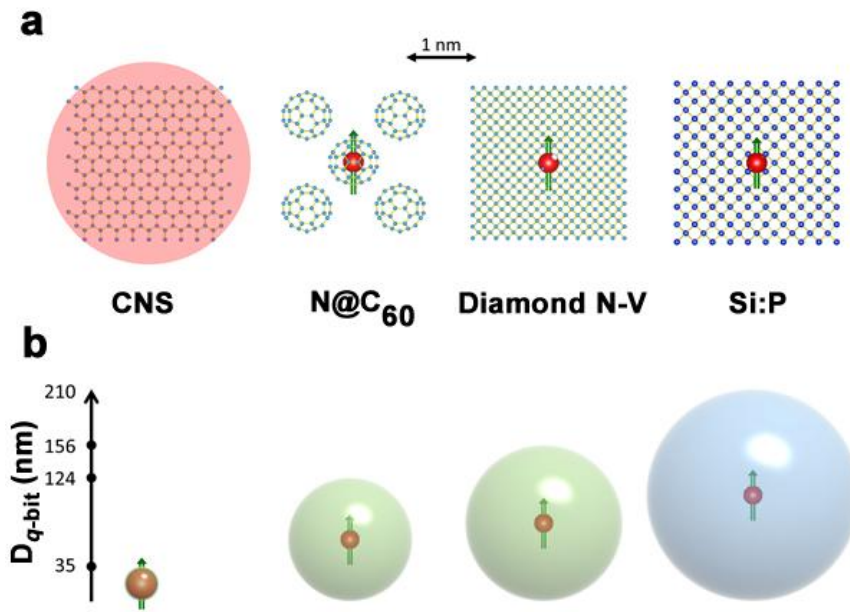

**Fig. 1. Comparison of itinerant and localised electron based qubits. a** 2×2 nm region from the centre of each respective system structure with the red dots representing the localised spins, which are surrounded by spacer atoms. In the case of carbon nanospheres (CNS) the spin information is encoded by a delocalised electron spin which spreads over the entire 35 nm diameter (shaded area) making the system more robust against external magnetic field fluctuations and hyperfine interactions enforced by nuclear spins. Consequently, high qubit density can be achieved without enhanced decoherence. **b** The sphere diameters comparing the required volume for different types of qubits[14,25,34].

The structural and chemical properties of the carbon nanospheres are inherently key to the observation of the long $T_1=T_2$. Transmission electron microscopy (TEM) images show the extensive formation of spherical carbon spanning micron scales (Fig. 2a and Supplementary Fig. S1). The as-prepared carbon nanospheres are a conglomeration of spherical bodies and after sonication in suspension they could appear as individual particles (Fig. 2b). In comparison to other nanoparticle quantum dots used as qubits, e.g. Mn doped PbS[35], the carbon nanospheres are relatively uniform with a size distribution of 37 nm ± 7 nm estimated from TEM images (Supplementary Fig. S2).

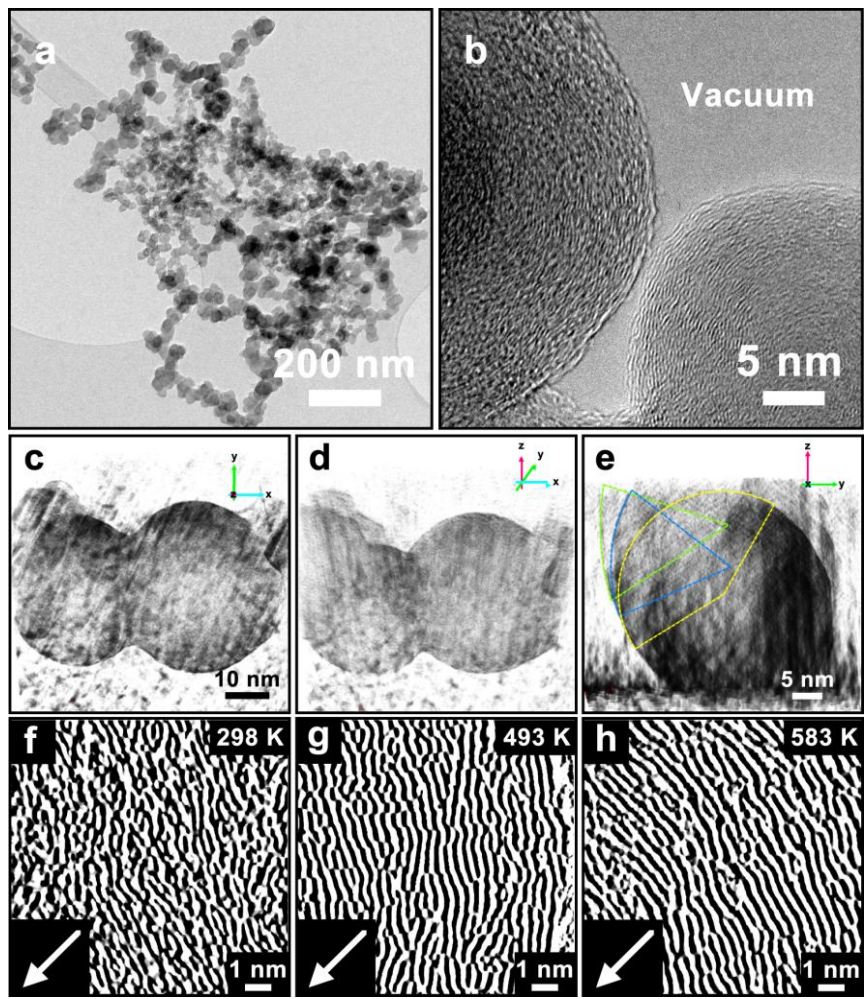

**Fig. 2. Carbon nanosphere structure and size. a** TEM image of carbon nanospheres on a carbon support. Darker regions of relative opacity are due to particle overlap. **b** High resolution TEM of discrete carbon nanospheres. **c-e** TEM 3-dimensional tomography reconstruction of the carbon nanospheres. The particles are spherical and also distort slightly to an ellipsoidal shape, with coalescence. Spheres appear transparent due to the high image contrast with the sputtered gold substrate, small sphere size, and extremely thin sphere layers. Scale bars in **d** were omitted for clarity. **f-h** In-situ variable temperature high resolution TEM images of the non-crystalline carbon nanosphere structure within regions of various spheres. The spheres remain non-crystalline

upon heating to 583 K. A high contrast is applied to the images to allow the graphite planes to be distinguished with a black outline. Arrows indicate direction towards the center of the sphere.

By means of TEM tomography a number of carbon nanospheres are also observed to have an asymmetric shape, which also results from the formation of joint graphitic layers of contacting particles (Figs. 2c–e), with the accretion of layers between nanospheres generally forming within a region of *ca.* 5 nm of the outer layers (Supplementary Fig. S3). High resolution TEM revealed the short graphitic fragments that comprise the carbon nanospheres are graphitic fragments that follow the curvature of a sphere, creating many open edges (Figs. 2f–h). The individual graphitic fragments in the carbon nanospheres are not curved and do not resemble the curvature in nanotubes or fullerenes. Rather, the fragments exhibit an intricate array of interplanar bonding all the way to the centre of the nanosphere even when heated to temperatures of 583 K. The carbon nanospheres are not hollow and show a continuation of the closed cage structure towards the centre.

X-ray photoelectron spectroscopy (XPS), thermogravimetric analysis (TGA), and Raman experiments were also performed and are found in Supplementary Figs. S4–S9. XPS indicated that the chemical structure is predominately conducting graphitic carbon (90.2 weight percent) containing surface bound oxygen (9.8 weight percent), and there was no inclusion of metals and other heavy atoms. TGA experiments confirmed the carbon nanosphere material did not contain residual precursor polyaromatic hydrocarbons and remained chemically and thermally stable even up to temperatures of 883 K. Valence band XPS revealed the presence of non-bonding $\pi$ and $\sigma$ orbitals as a result of fragmented sheets that contained carbon arranged in a distorted hexagonal network.

The carbon nanospheres are easily synthesised[36] and readily processable (Fig. 3), yielding a homogenous material that is structurally highly non-crystalline (Fig. 2). Thus the carbon nanospheres can be reliably employed for spintronics applications with minimal processing and without the need for fabricating a well-defined crystal structure to achieve long $T_1$ and $T_2$. Furthermore, in a conducting carbon nanosphere qubit system, the rich chemistry of carbon can, in future experiments, allow for a myriad of non-covalent and covalent interactions to connect the nanospheres to conducting electronic device surfaces[37,38]. Finally, the carbon nanospheres are of a size that can be isolated on a surface from the 'top-down' using micromanipulator probe tips (Fig. 3d), and this in future experiments can allow for building qubit ensembles[39].

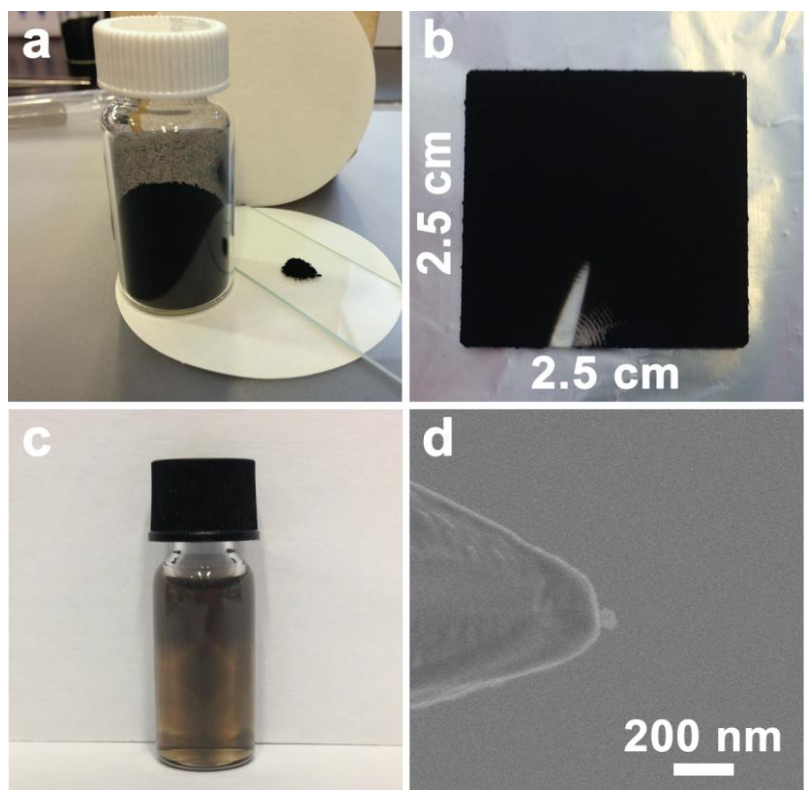

**Fig. 3. Sample processing of carbon nanospheres. a** 300 mg of carbon nanospheres collected as a solid powder in a sample tube and 10 mg on a glass slide, stable in air. **b** Carbon nanospheres directly deposited during synthesis onto quartz. **c** A small amount of carbon nanospheres dispersed in 2 ml of ethanol by sonication. **d** SEM image of an individual ~50 mn CNS physically positioned on a Si substrate using a 200 nm tungsten manipulator tip.

Here we develop in more details our key observations obtained by ESR. At 300 K and at 9.4 GHz frequency the continuous-wave ESR line width (peak-to-peak) is $\Delta H$=0.056 mT (inset Fig. 4a and Fig. S10) and the $g$-value is 2.00225 (inset Fig. 4b). This is a remarkably narrow conduction-electron spin ESR line testifying the long spin relaxation times. The observed spectra had, to a high precision, homogeneously broadened Lorentzian line-shapes and the deviation from the Lorentzian line-shape in the entire spectra was less than 5%, (see Figure S10 as an example), which reveals the itinerant nature of the spins. The observed line width determined by continuous-wave ESR is identical within experimental error with the $T_2$ derived Lorentzian width. Note that the size distribution of the carbon nanospheres has a negligible effect on the linewidth at 9.4 GHz because of the motional narrowing of conduction electrons (see Supplementary Fig. S11). The hyperfine ESR lines of $^{13}$C were also absent, (which are readily observable for localized spins[14,25]) due to motional narrowing of conduction electrons[26]. The $g$-factor is characteristic to conduction electrons of carbon, and it does not originate from metallic inclusions (in agreement with our chemical analysis) or from localised paramagnetic 'dangling' bonds of carbon (commonly with $g$=2.00282)[40].

In addition to the continuous-wave ESR experiment, where detection and spin rotation occur at the same time, we extended our experiments to probe the spin relaxation dynamics of $T_1$ and $T_2$ independently using pulsed ESR (Fig. 4a). At 9.5 GHz frequency and 300 K, with good approximation we found that the intrinsic $T_1$=$T_2$=175 ns. Pulsed ESR therefore simultaneously validated our continuous-wave ESR results and verified that the line-shape obtained by continuous-wave ESR was indeed homogenous as expected for itinerant electrons.

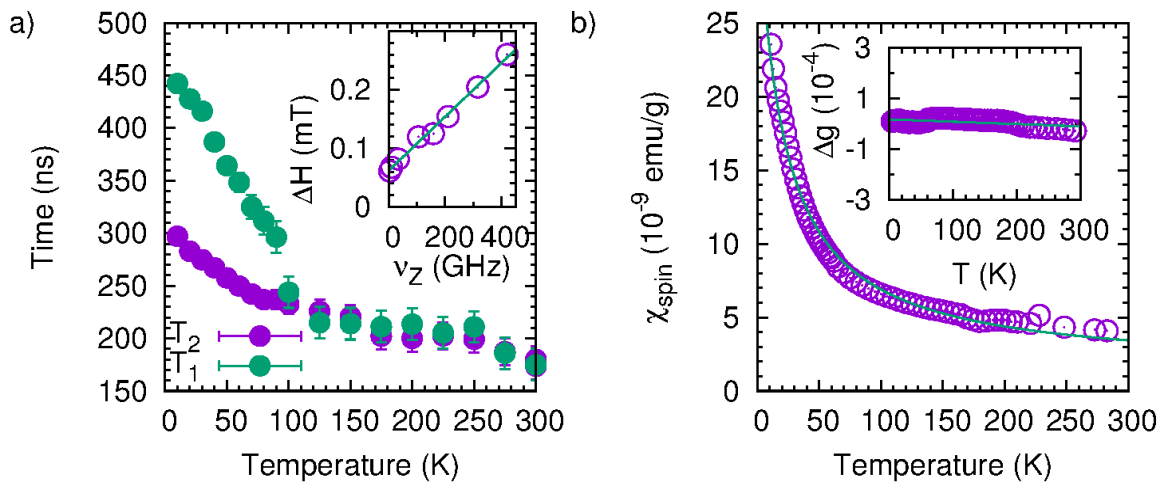

Figure 4. **Characterization of the spin system of the carbon spheres by ESR. a** Temperature dependence of $T_1$ and $T_2$ at $v_Z$=9.5 GHz. The inset shows the ESR linewidth plotted as a function of the Zeeman energy, $E_Z$=$hv_Z$ measured by a multi-frequency ESR at 300 K.: The linear fit (straight solid line) using equation 1, with $T_1$=$T_2$=175 ns gives δ=1 meV. **b** Temperature dependence of the spin susceptibility, $\chi_{spin}$, measured by ESR at $v_Z$=9.4 GHz with an overlaying Curie-Weiss line, characteristic to small paramagnetic particles. Inset in **b** shows the temperature independent $g$-factor shift Δ$g$ relative to the free electron $g$-value, in good agreement with a material exhibiting very weak spin orbit coupling.

The temperature dependent properties of the ESR spectra (Fig. 4 and see also Supplementary Fig. S12) support that conduction electrons are confined within the carbon nanospheres. The $g$-factor was temperature independent (inset Fig. 4b), which is in good agreement with general observations in metals with weak spin orbit coupling and in graphitic nanoparticles[41]. The resulting spin susceptibility is temperature dependent, following a Curie-Weiss dependence, as one may expect for nanoparticles of metals (Figure 4b)[21,42].

Multi-frequency ESR in the 4–420 GHz frequency and 2–300 K temperature range also confirmed that conduction electrons are confined within the carbon nanoparticles (inset Fig. 4b and Fig. S12)[43,44]. The ESR linewidth revealed a linear increase with increasing magnetic field at 300 K (Fig. 4a inset and see also Supplementary Fig. S12). Note that in the case of bulk metals $\Delta H$ is solely determined by spin orbit coupling thus it is independent of the magnetic field[21,45]. However, the behaviour observed when the carbon nanospheres experience a variation in external magnetic field is characteristic to conduction electrons enclosed in nanoparticles where $T_1$ and $T_2$ are determined by both the spin-orbit interaction and electron confinement[21]. This broadening of $\Delta H$ for itinerant electrons confined on small particles follows[46]:

$$\Delta H = E_Z/(\delta \gamma_e T_2) \qquad (1)$$

where $E_Z = h\nu_Z$ is the Zeeman energy, $\delta$ is the average electronic energy level spacing, and $\gamma_e$ is the electron gyromagnetic ratio[46]. Using the measured $T_1=T_2=175$ ns we can extract a $\delta=1$ meV for the average electronic energy level spacing (linear line of best fit in Fig. 4a inset). From this value one can calculate back the size of the carbon spheres by following Kubo's calculations for a small, almost spherical, metallic particle[42]:

$$\delta = 4m_e v_F^2/3n\pi L^3 \qquad (2)$$

where $m_e$ is the free electron mass, $v_F=10^6$ m/s is a typical Fermi velocity for graphene, and $n=2.3$ g/cm$^3$ is the atomic density of the carbon nanospheres. These values yield an effective linear particle size of $L=40$ nm. This particle size is in good agreement with that obtained from TEM images. Furthermore, the increase in $T_1$ and $T_2$ as the temperature is decreased (Fig. 4a and see also Supplementary Fig. S12) is characteristic to metals[21,41,45], where electron-phonon scattering due to spin-orbit interaction is responsible for the temperature induced shortening of $T_1$ and $T_2$[42,45,46].

As the temperature was decreased, $T_2$ reached 300 ns at 4 K while $T_1$ reached 450 ns (Fig. 4a). There is a deviation from the $T_1=T_2$ dependence below ~100 K. During the delineation of $T_1$ and $T_2$ below ~100 K, $T_1$ and $T_2$ nevertheless continue to increase at different rates. The electron spin dynamics of $T_1$ is directly related to phonon dampening in disordered graphitic sheets[26,41] and the existence of Wallis-type[47] local phonon modes. Future studies have to show if further localisation at specific defect sites occurs at very low temperatures. As the temperature is increased, the calculations of Andersson *et al.*[41] indicate that the observed shortening of $T_1$ and $T_2$ is caused by the scattering of conduction electrons by the potentials of peripheral atoms having edge-inherited electronic and lattice dynamical features and the excitation of low energy phonons.

From our structural, chemical and electronic characterisation we propose that the non-bonding orbitals associated to the structural imperfections induce conduction electrons to the system and significantly enhance the electron density of states. This is in agreement with theoretical works[41] that predict the presence of an additional band superimposed upon the bonding $\pi$ and the anti-bonding $\pi^*$ bands around the Fermi energy in nanometre size disordered graphitic fragments, and our observations of the changes in the $p_z$ wave functions in the $p$-$\pi$ band evolutions with temperature near the Fermi energy level (see Supplementary Fig. S5).

The carbon nanospheres also contain covalently bonded oxygen that contribute to the disorder within the graphitic lattices (Figure 3f–3h and Supplementary Table S1 and Figure S9). We note that the removal of oxygen groups by thermal decomposition introduces non-bonding $\pi$ and $\sigma$ orbitals (Supplementary Figs. S4 and S5). In future experiments, greater robustness against spin decoherence may be achieved by the removal of adatoms to enhance the electronic density of states[32].

Although the carbon nanospheres are highly defective, the intrinsically weak spin-orbit interaction of carbon has allowed for long $T_1$ and $T_2$ to persist even at 300 K. In nanotubes and fullerenes an increase in spin orbit coupling due to graphene sheet curvature may contribute to the shortening of $T_1$ and $T_2$[27,28], however the individual graphitic flakes in the carbon nanospheres are not curved. The $T_1$=$T_2$ in the carbon nanospheres is remarkably a two orders of magnitude enhancement over that found in conducting crystalline graphene[31]. We attribute part of this increase in $T_1$ and $T_2$ to quantum confinement effects readily observed in conducting nanostructures[4,6,7,21].

Following our observations of magnetically induced spin polarization, we coherently rotated the electron spin at 300 K by applying microwave power bursts of increasing duration and with variable power (Fig. 5a). We observed that the magnetization of the carbon nanospheres oscillates periodically with pulse duration, the oscillation frequency being proportional to the square root of the microwave power. This oscillation indicates that we have performed deliberate and coherent electron spin rotations (driving of electron spins between two Zeeman-split energy levels), or Rabi oscillations. This is the sign that one can manipulate spins both for spintronics and quantum information processing (see also Supplementary Information Methods)[8,48]. Fourier-transformed Rabi oscillation signals show a single component characteristic to electron spin-1/2 qubit (Fig. 5b). A key characteristic of the Rabi process is a linear dependence of the Rabi frequency on the microwave field strength $\mathbf{B_1}$ ($f_{Rabi} = g\mu_B B_1/h$). We verify this by extracting the Rabi frequency from a fit of the Fourier-transformed signal of Fig. 5b with a Gaussian line, which gives the expected linear behaviour that is proportional to $\mathbf{B_1}$ (Fig. 5c). This demonstrated the capacity to rotate the spin qubits on the carbon nanospheres arbitrarily to any point on the Bloch sphere[10] at temperatures as high as 300 K. Rabi oscillations can be observed for approximately 400 ns, consistent with decoherence by $T_2$. The increase in linewidth of the Fourier transformed oscillations with microwave power (see Fig. 5b) is caused by microwave field inhomogeneity.

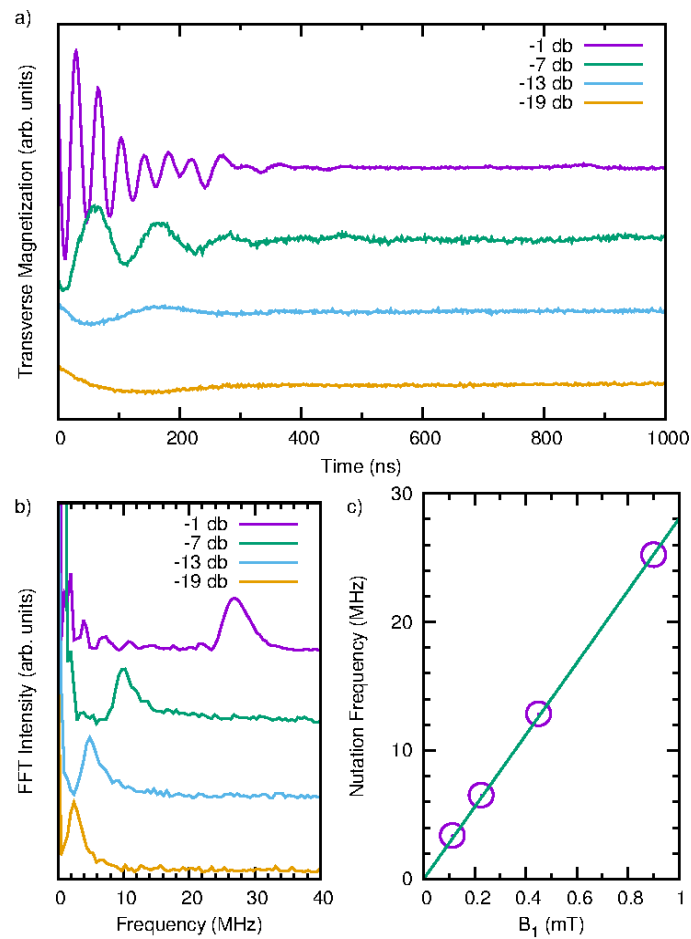

Fig. 5. **Spin control over electron qubits confined to carbon nanospheres at 300 K. a)** Rabi oscillations of the electron qubits at 300 K and $B_0$=337 mT for different microwave powers. **b)** Fourier transform of the Rabi oscillations, with the signal shifted vertically for clarity. **c)** Rabi frequency is proportional with the square root of the power. The value 25 MHz observed for the maximum power is consistent with the previously determined maximum $B_1$ field in the dielectric cavity of 0.9 mT, indicating the presence of an effective spin $S = 1/2$.

We have demonstrated that a long itinerant electron spin lifetime in a magnetically homogenous conducting material can be achieved at room temperature. Through the controlled polarization of coherent electron spins in these carbon nanospheres, we have showed that this electron spin lifetime exceeds the prerequisite for applications in spintronics and quantum information processing. This is possible through electron confinement to nanometre sized, non-crystalline yet metal-like carbon spheres. This work effectively bridges the disparate research directions in the fields of inorganic and molecular materials for electron spin qubits and has broad applicability: spin qubits can now be manipulated at room temperature without the need for isotopically engineering a host material, diluting the spin-carrying molecule, cryogenic temperatures, the preparation of well-defined crystal structures, or the use of metals. The facile preparation of a carbon material using common laboratory reagents, combined with the use of well-established electron spin manipulation measurements at room temperature, effectively reduces many of the technological barriers to realising practical quantum computing and spintronics using solid state materials.

The spin manipulation experiments described here were performed on a large number of carbon nanospheres. Although the material can be readily chemically processed, it is prepared in a form suitable for device processing: we have demonstrated that the conducting nanospheres can be isolated on a silicon surface by physically manipulating individual nanospheres. In principal, this may provide an initial avenue to high density qubits arrays of nanospheres that are integrated onto existing silicon technologies or thin-film based electronics.

**Methods**

**Sample Preparation.** The preparation of the carbon nanospheres is described in detail elsewhere[36] and can be summarized as the soot product resulting from the partial combustion of naphthalene in air, which is collected and heated at 473 K under dynamic vacuum for 72 h. Approximately 100 mg was prepared for all experimental procedures.

**Transmission electron microscopy.** Samples were analysed using a field emission JEOL3000F operated at 300 kV. Particle size distribution and topography image analysis was performed using freely available ImageJ 1.48v software (http://imagej.nih.gov/ij). Energy filtered tomography images were obtained using a JEOL JEM 2200FS Field Emission Microscope operated at 200 kV using a high tilt holder, with an in-column omega filter, and objective aperture applied. JEOL recorder software v2.48.1.1 was used to collect the topography images. JEOL Composer and JEOL Visualizer-kai programs were used to reconstruct the tomography images.

**X-ray photoelectron spectroscopy.** Measurements were conducted using an ESCALAB250Xi instrument manufactured by Thermo Scientific, UK. The background vacuum was better than $2\times10^{-9}$ mbar. A monochromated Al $K_\alpha$ (energy hv=1486.68 eV) was used with a spot size of 500 μm. The fine carbon powder was manually pressed onto indium foil for analysis or pressed into a disc prior to *in-situ* heating experiments. Curve fitting was performed using the Scienta ESCA300 data-system software. Binding energy reference was C $1s$ = 285.0 eV for adventitious carbon.

**Continuous wave electron spin resonance.** Experiments were performed on a home built quasi-optical spectrometer operated in the 55–420 GHz frequency range in a corresponding 0–16 T field range[43,44]. At low frequencies at 4, 9.4 and 34 GHz a Bruker elexsys E500 spectrometer was used. For a typical experiment about 1 mg of the carbon sample was weighed and then sealed in a quartz ESR tube after being heated at 500 K under dynamic vacuum overnight. For temperature and frequency dependent experiments the magnetic field modulation amplitude was smaller than 0.01 mT and the microwave power was set to 0.2 μW to avoid signal distortion. For *g*-factor reference a polycrystalline $KC_{60}$ powder was used with g=2.0006.

**Pulsed electron spin resonance.** Experiments were performed at 9.4 and 34 GHz using Bruker ElexSys 580 and 680 spectrometers. For $T_2$ determination we used a simple 2 pulse sequence, invoking a first π/2 pulse of 16 ns, and an initial delay of the second pulse of 300 ns. The resulting echo was integrated over 175 ns. For $T_1$ a 3-pulse sequence was used. An initial π pulse of 32 ns was followed after 300 ns delay by a 2-pulse echo sequence with 200 ns initial pulse delay for monitoring the inversion recovery. For Rabi experiments a single π/2 pulse was used. Its length was incremented by 2 or 4 ns. After a delay of approximately 84 ns with respect to the pulse ending the signal was observed with a short integrating time of 16 ns.

**Acknowledgments**

Financial support of Swiss National Science Foundation is acknowledged. M.C. acknowledges financial support from The University of Sydney.

**Author Contributions**

M.C. synthesised the materials. M.C. and B.N. and K-P.D derived the experimental results. All authors discussed the results and contributed to the writing of the manuscript.

**Supplementary Information**

Further description of materials and methods, TEM, SEM, XPS, TGA, Raman spectroscopy, and ESR data available.

Supplementary Information

# Spin manipulation in metal-like carbon nanospheres at room temperature

Bálint Náfrádi[1*], Mohammad Choucair[2*], Klaus-Peter Dinse[3], László Forró[1]

[1]Institute of Condensed Matter Physics, École Polytechnique Fédérale de Lausanne, Lausanne 1015, Switzerland.

[2]School of Chemistry, University of Sydney, Sydney 2006, Australia.

[3]Institut für Experimentalphysik, Freie Universität Berlin, Arnimallee 14, 14195 Berlin, Germany.

[*]Correspondence to: balint.nafradi@epfl.ch, mohammad.choucair@sydney.edu.au.

## Supplementary Information – Methods

For transmission electron microscopy (TEM) samples were analysed on a field emission JEOL3000F operated at 300 kV. Scanning electron microscopy (SEM) was performed using a Zeiss Ultra Plus.

X-ray photoelectron spectroscopy (XPS) measurements were conducted using an ESCALAB250Xi instrument manufactured by Thermo Scientific, UK. The background vacuum was better than $2\times10^{-9}$ mbar. A monochromated Al $K_\alpha$ (energy hν=1486.68 eV) was used with a spot size of 500 μm. The fine carbon powder was manually pressed onto indium foil for analysis or pressed into a disc prior to *in-situ* heating experiments. Curve fitting was performed using the Scienta ESCA300 data-system software. Binding energy reference was C 1$s$ = 285.0 eV for adventitious carbon.

Raman spectroscopy was performed using Argon 514 nm excitation laser on a Renishaw Raman inVia Reflex with a notch and edge filter cut-off of 100 cm$^{-1}$.

Modulated Thermogravimetric analysis measurements (mTGA) were obtained using a TA HiRes Discovery TGA in Modulated TGA Mode with a heating profile of 2 °C/min under the flow of 20 mL/min high purity $N_2$ with sinusoidal temperature amplitude of 4 °C and period of 200 s. A 100 μL alumina pan was used. Evolved gas analysis (TGA GC-MS) was performed using a Perkin Elmer Thermogravimetric Analyzer Pyris 1 coupled to a Perkin Elmer Gas Chromatograph Clarus® 680 and Mass Spectrometer Clarus® SQ 8 C.

ESR experiments revealed a presence of a single narrow ($\Delta H$=0.05 mT) Lorentzian line with $g$=2.00225 at 9.4 GHz frequency (Fig. S10). The spectral resolution of ESR is proportional to the frequency. The deviations from the Lorentzian shape even at 420 GHz were smaller than

5%. There was no *g*-factor anisotropy observed within the resolution of the 420 GHz measurements of $\Delta g < 10^{-6}$. Note that in carbon the *g*-factor values of localized paramagnetic centres are in the 2.0025-2.0050 range with anisotropies $\Delta g$ in the order of $\sim 5 \times 10^{-4}$.[1]

In the Zeeman effect a uniform external magnetic field is applied to a collection of atoms, and measurements are made of the potential energies of orientation in the field of the magnetic dipole moments of the atoms. Except when an atom is in the $^1S_0$ state, it will have a total magnetic dipole moment, **μ**, due to the orbital (**μ**$_l$) and spin (**μ**$_s$) magnetic dipole moments of the optically active electrons. When this magnetic dipole moment is in an external magnetic field **B** it will have a potential energy of orientation[2]:

$$\Delta E = - \mathbf{\mu} \cdot \mathbf{B} \qquad (1)$$

Each energy level will be split into several discrete components corresponding to the various values of $\Delta E$ associated with the different quantised orientations of **μ** relative to the direction of **B**. Simply put: because the atom has a magnetic dipole moment the energy of the atom depends upon which of the possible orientations it assumes in the external magnetic field. Due to the quantised selection rules the total spin angular momentum and magnetic dipole moment change orientation in an atomic transition.

Now if we consider a spin-½ qubit system between two Zeeman-split energy levels (*i.e.* +1/2 and -1/2) in an oscillatory external field, the probability of the qubit being in the excited energy level at a given time *t* is found from the Bloch equations[3] and given by[4]:

$$P(t) = (\omega_1/\Omega)^2 \sin^2(\Omega t/2) \qquad (2)$$

where $\Omega = \sqrt{(\omega - \omega_0)^2 + \omega_1^2}$ ; and $\omega_0 = \gamma B_0$, $\omega_1 = \gamma B_1$, and $\gamma$ is the gyromagnetic ratio.

This is called Rabi oscillation; this shows that there is a finite, oscillatory, probability of finding the system in and excited state when the system is originally in the ground state. The maximum amplitude for oscillation is achieved at frequency $\omega = \omega_0$ which is the condition for resonance. At resonance, the transition probability is then given by:

$$P(t) = \sin^2(\omega_1 t/2) \qquad (3)$$

Where $\omega_1$ is the Rabi frequency. Thus the detected Rabi oscillations of several cycles can be taken as evidence that the system allows the deliberate preparation of any superposition of a two level spin-½ system. For example, to go from one state (ground state) to the other (excited state) we can adjust the time *t* during which the rotating field acts such that $\omega_1 t/2 = \pi/2$ (*i.e.* $t = \pi/\omega_1$); this is called a $\pi$ pulse. If a time intermediate between 0 and $\pi/\omega_1$ is chosen, *e.g.* in the case for $t = \pi/2\omega_1$, we obtain a $\pi/2$ pulse, and this results in a superposition of both states.

## Supplementary Information – Electron Microscopy

SEM images show an extensive formation of carbon nanoparticles spanning micron scales, Figure S1. The disordered non-crystalline carbon structure was observed to be maintained during heating, Figure S3. Due to carbon decomposition on the carbon nanospheres (CNSs) during heating in the TEM, continuous observation of the layers within the same CNSs was not always possible and hence a number of different CNSs were investigated.

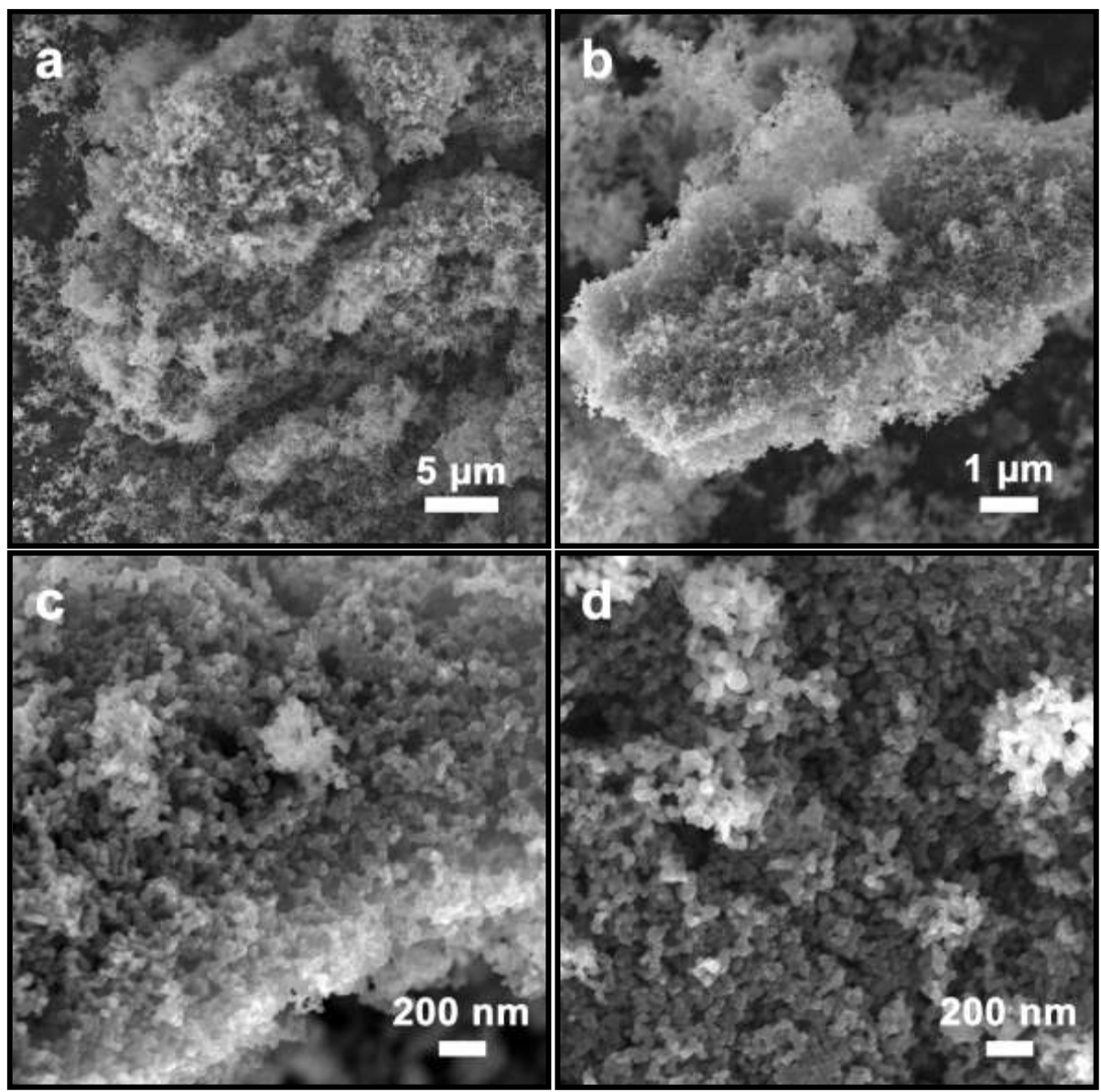

**Figure S1. SEM images.** (a) and (b) of the CNSs at low magnification, and (c) and (d) at higher magnification, with (c) being from a region in (b).

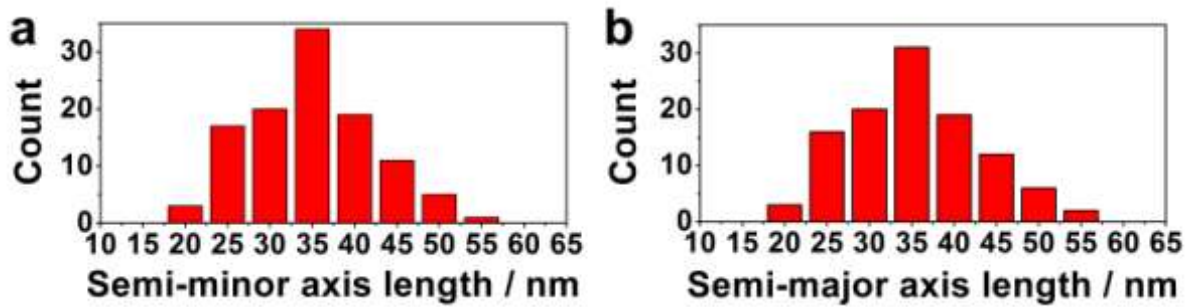

**Figure S2. Carbon nanosphere particle size distribution.** (a) and (b) Narrow particle size distribution histograms (35 nm range) obtained from the TEM image in Figure 2a assuming ellipsoidal particles with semi-major and semi-minor axis lengths plotted respectively.

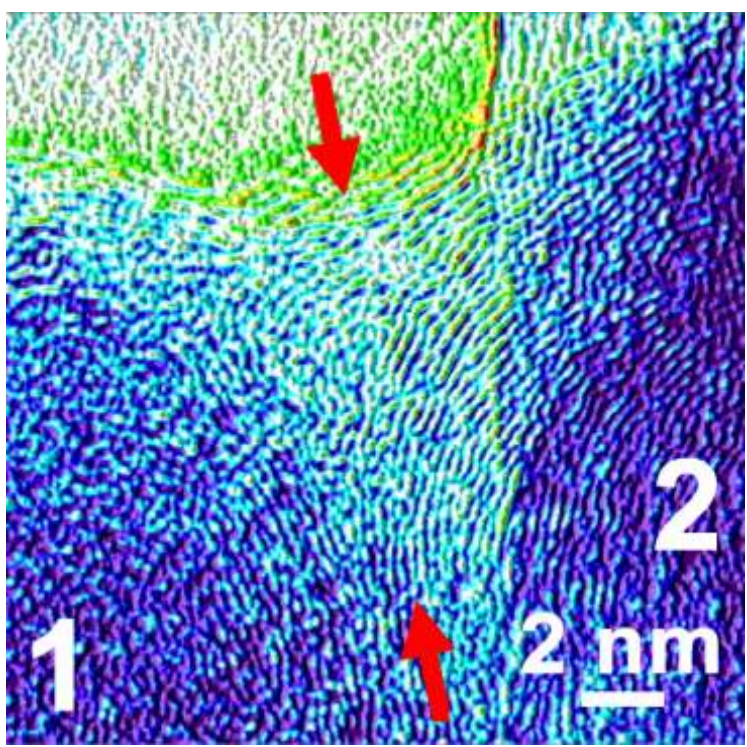

**Figure S3. Carbon nanosphere accretion.** A high resolution TEM of two nanospheres (1 and 2) shown as a colour-surface topography with darker regions indicating shell overlap and the region of coalescence between them indicated by the arrows. Sphere 1 is suspended over a vacuum while 2 is on a carbon support, with ca. 15 outer layers coalescent.

# Supplementary Information – X-ray Photoelectron Spectroscopy

The main core C 1$s$ envelope was representative of an asymmetric peak commonly obtained for conducting graphitic materials; having a low level of oxidation and a very narrow peak width at half-maximum (FHWM less than 1.2 eV) and positioned at a binding energy corresponding to pure graphitic material 284.5 eV (Figure S4a)[5,6]. Upon heating, the C 1$s$ peak did not shift from the $sp^2$ graphite binding energy position of 284.5 eV or change notably in width at half maximum.

The total atomic ratio at 298 K of carbon to oxygen was *ca.* 12:1 which increased to *ca.* 14:1 as the material was heated, Figure S4. The increase in temperature resulted in the removal of oxygen in the form of $CO_2$ (Figure S8). Curve fitting employed for the O 1$s$ line indicated both O=C (533.2 eV) and O–C (531.8 eV) chemical bonding environments were present in the onion-like carbon nanospheres. It was inappropriate to quantify the individual oxygen environments present as there was low total oxygen content (less than 10 at.%), a small difference in oxygen lost during heating (less than 0.8 at.%), and a broad O 1$s$ line peak (FHWM *ca.* 3.8 eV), Figure S4b. The XPS results indicated that the CNSs remained chemically and thermally stable even up to temperatures of 583 K.

The valence band XPS spectrum (Figure S5) shows a fairly broad, intense peak located between 16 and 23 eV, a narrower less intense peak with a well-defined minimum located between 12 to 15 eV (both assigned to C 2$s$), and a very broad and decidedly weaker structure tailing off and extending from 12 eV to the cut-off energy ($p$-σ peak) typical of graphitic material.[7-9] The C 2$s$ peak had two peaks (10–25 eV) which strongly suggested the presence of an $sp^2$ network made up of six-fold rings, as this feature is known to 'wash-out' by the presence of an increased number odd-membered rings in a random network.[8] A single O 2$s$ contribution is also observed between 24 to 29 eV.[9] The positions of the band peaks do not change with temperature. The $p$-π states are not apparent in the as prepared CNSs and only appear as a shoulder on the leading edge of $p$-σ peak after annealing beyond 363 K and persists even when cooling to 123 K. The emergence of the $p$-π band may arise from changes in the $p_z$ wave functions at large radii due to the delocalised nature of the $p$-π orbitals.[7] The evolution of $p$-π states with temperature closely resembled that observed for amorphous/non-crystalline carbon (above 623 K in amorphous carbon)[8], and the presence of an O 2$s$ contribution was similar to that observed in partially oxidised graphitic fibres[9].

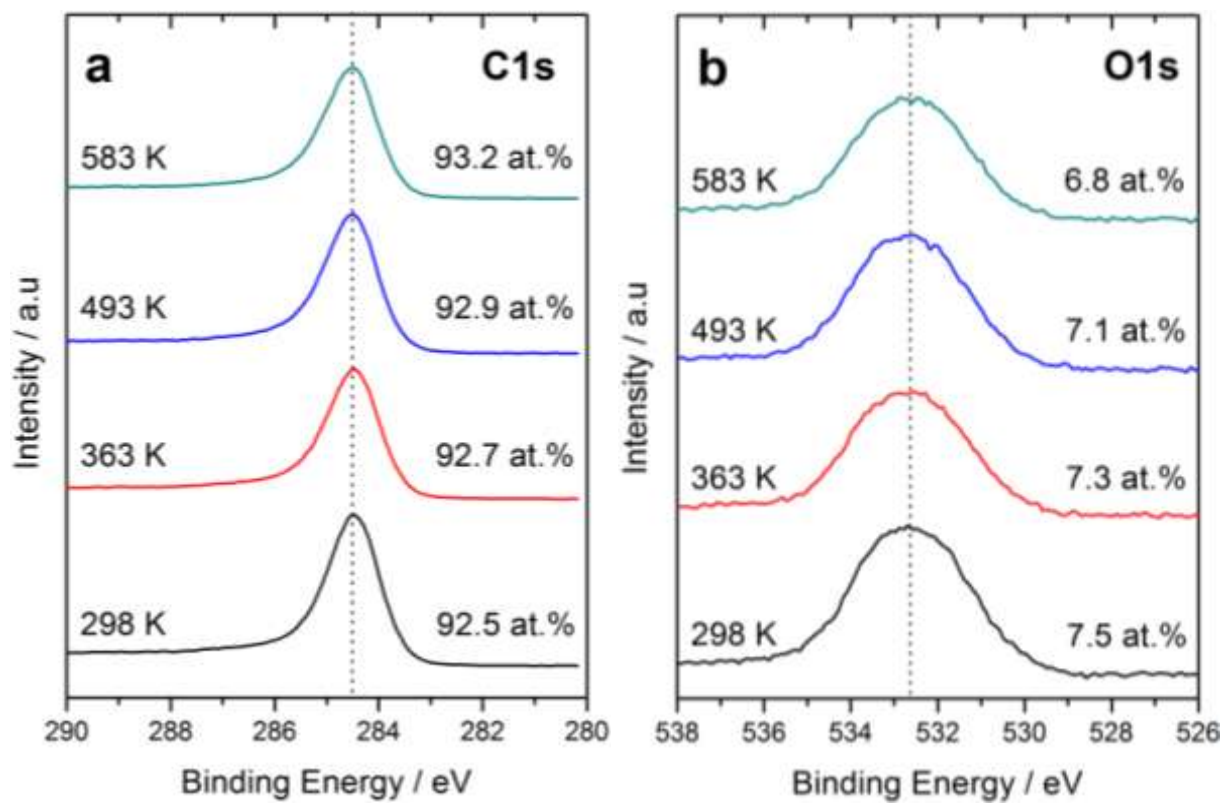

**Figure S4. XPS spectra with corresponding atomic percentage contributions of the core C 1s and O 1s lines during *in-situ* heating.** Dashed line in (a) is at 284.5 eV and in (b) 532.6 eV.

**Table S1.** $sp^2$ to $sp^3$ carbon content in the sample obtained from the XPS C 1s core line.

| Temperature / K | $sp^2$ / % | $sp^3$ / % |
|---|---|---|
| 298 | 66 | 34 |
| 363 | 62 | 38 |
| 493 | 60 | 40 |
| 583 | 60 | 40 |

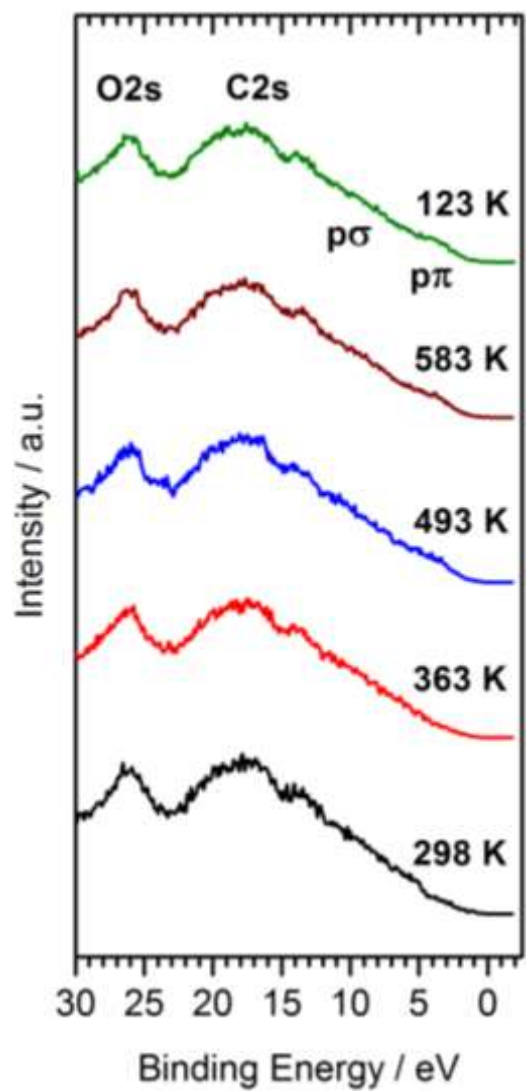

**Figure S5.** *In-situ* valence band XPS spectra of CNSs at various temperatures. The material is first heated to 593 K then cooled to 123 K.

## Supplementary Information– Thermo-gravimetric Analysis

In Figure S6 two weight loss events occurred very distinctly, at 583 K and 883 K, attributed to the removal of chemically bound oxygen in the form of $CO_2$, (see Figures S4 and S7). Less than 2 weight percent (wt.%) was lost at 493 K (attributed to degassing, removal of adsorbed $H_2O$), less than 5 wt.% at 583 K, and remarkably only less than 10 wt.% at 923 K. The large activation energies associated with the main weight loss events (300–400 kJ/mol) were evidence of very slow decomposition of the carbon, Figure S6. However, it remains unclear whether the removal of oxygen from O–C and O=C groups occurs separately.

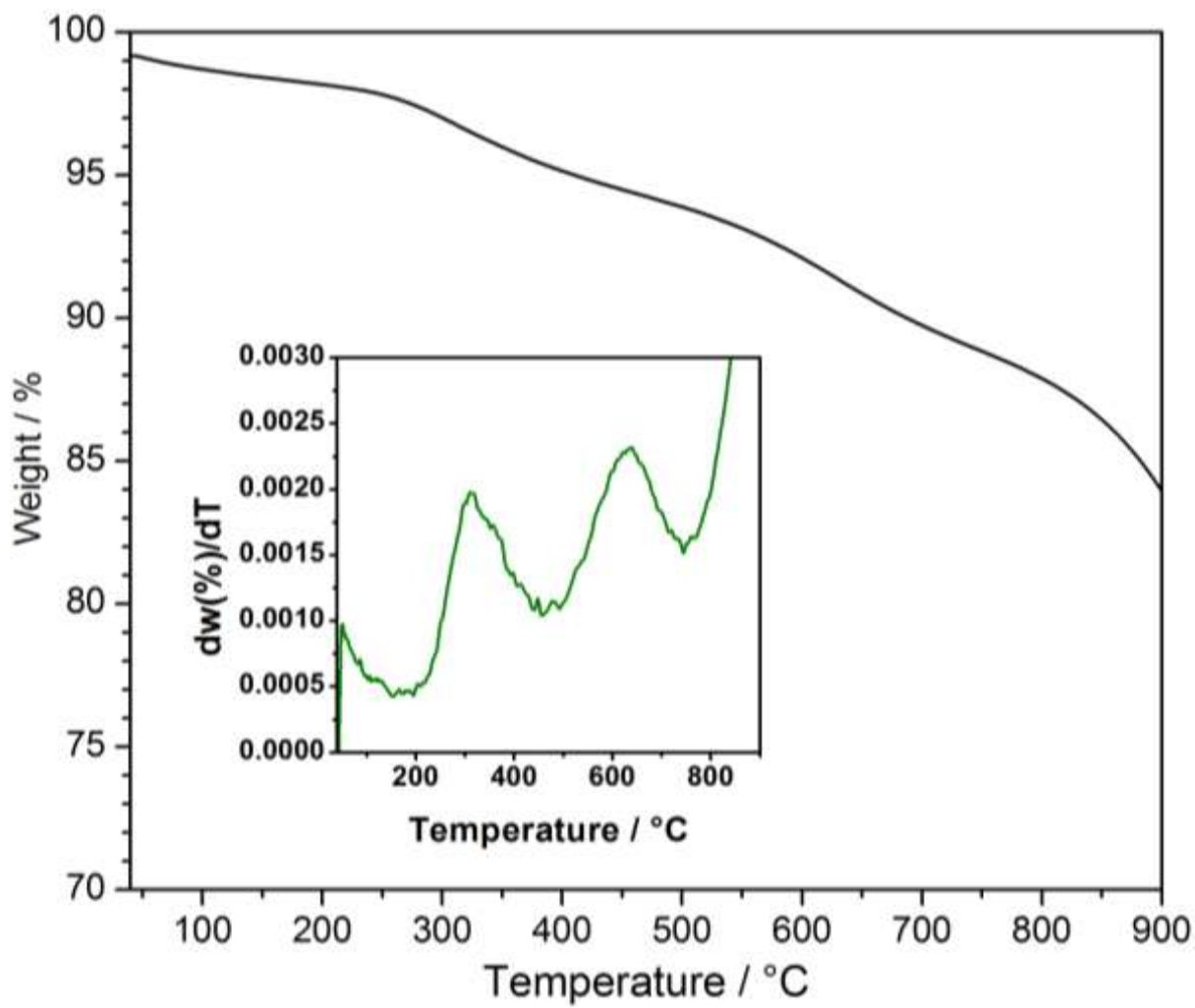

**Figure S6. High resolution TGA of the CNSs material.** Inset shows the derivative of weight loss as a function of temperature with 2 prominent weight loss features at *ca.* 310°C and 610°C. Less than 2% of weight loss event occurred which can be attributed to degas and solvent losses (temperature up to 150°C).

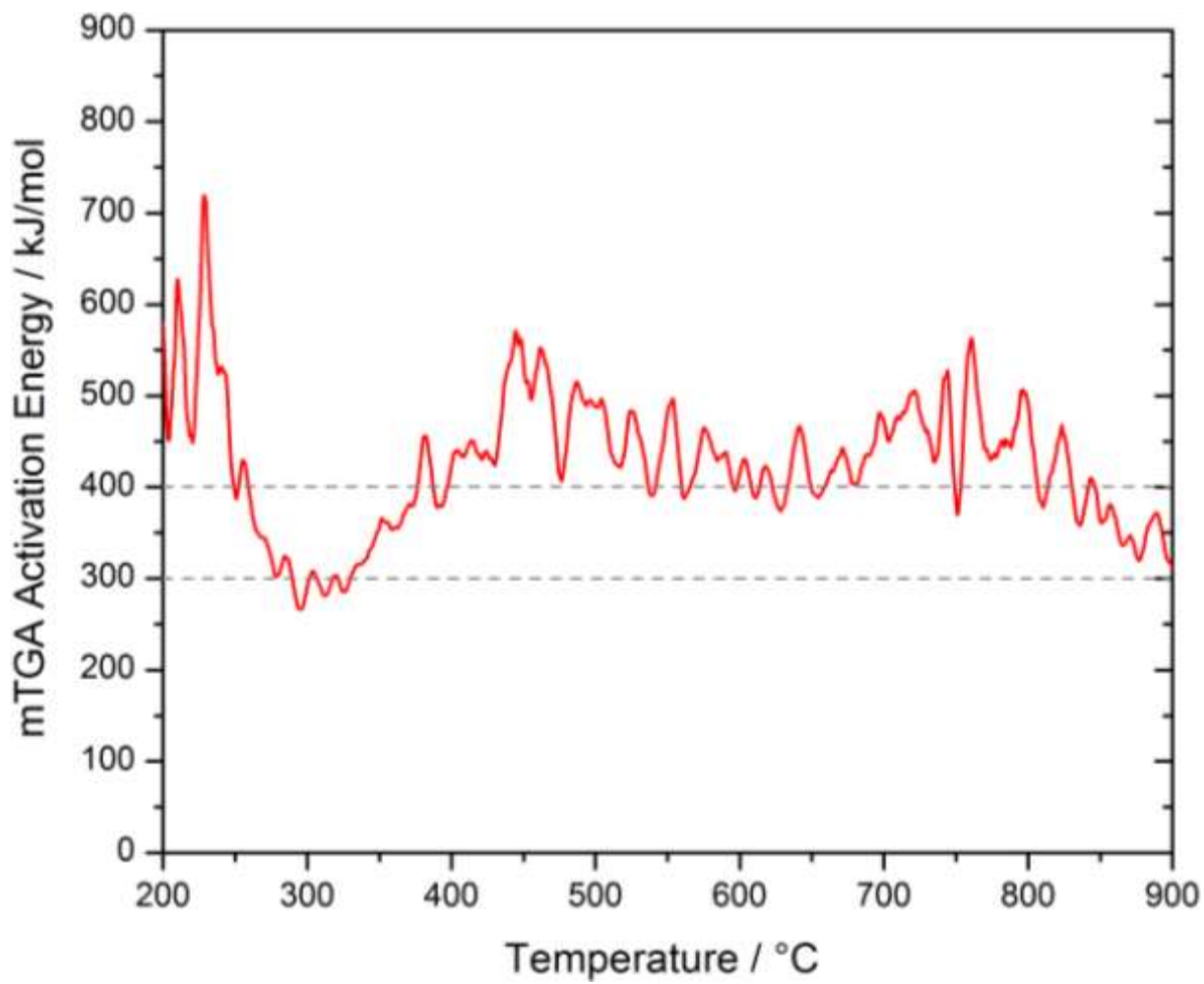

**Figure S7. Activation energy during heating to 900°C obtained from the modulated TGA experiment on the CNSs.** The corresponding major weight loss events at *ca.* 310°C and 610°C have activation energies of *ca.* 300 kJ/mol and 400 kJ/mol respectively, indicating very slow kinetics of decomposition.

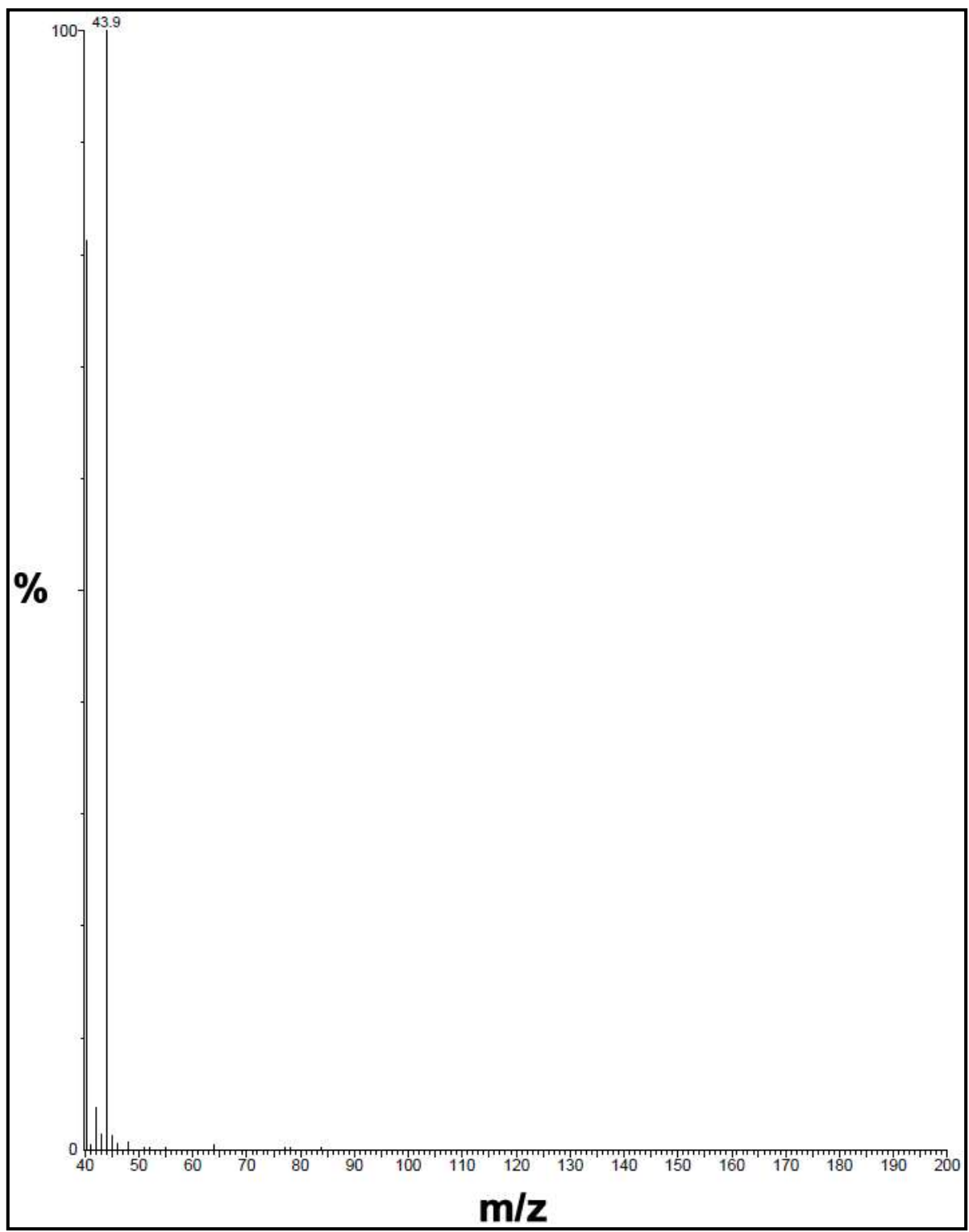

**Figure S8. Typical TGA-mass spectra**. Heating of CNSs from 150°C to 610°C, showing a $m/z$ corresponding to $CO_2$. No evidence of naphthalene or other polyaromatic hydrocarbons in the sample.

**Supplementary Information – Raman Spectroscopy**

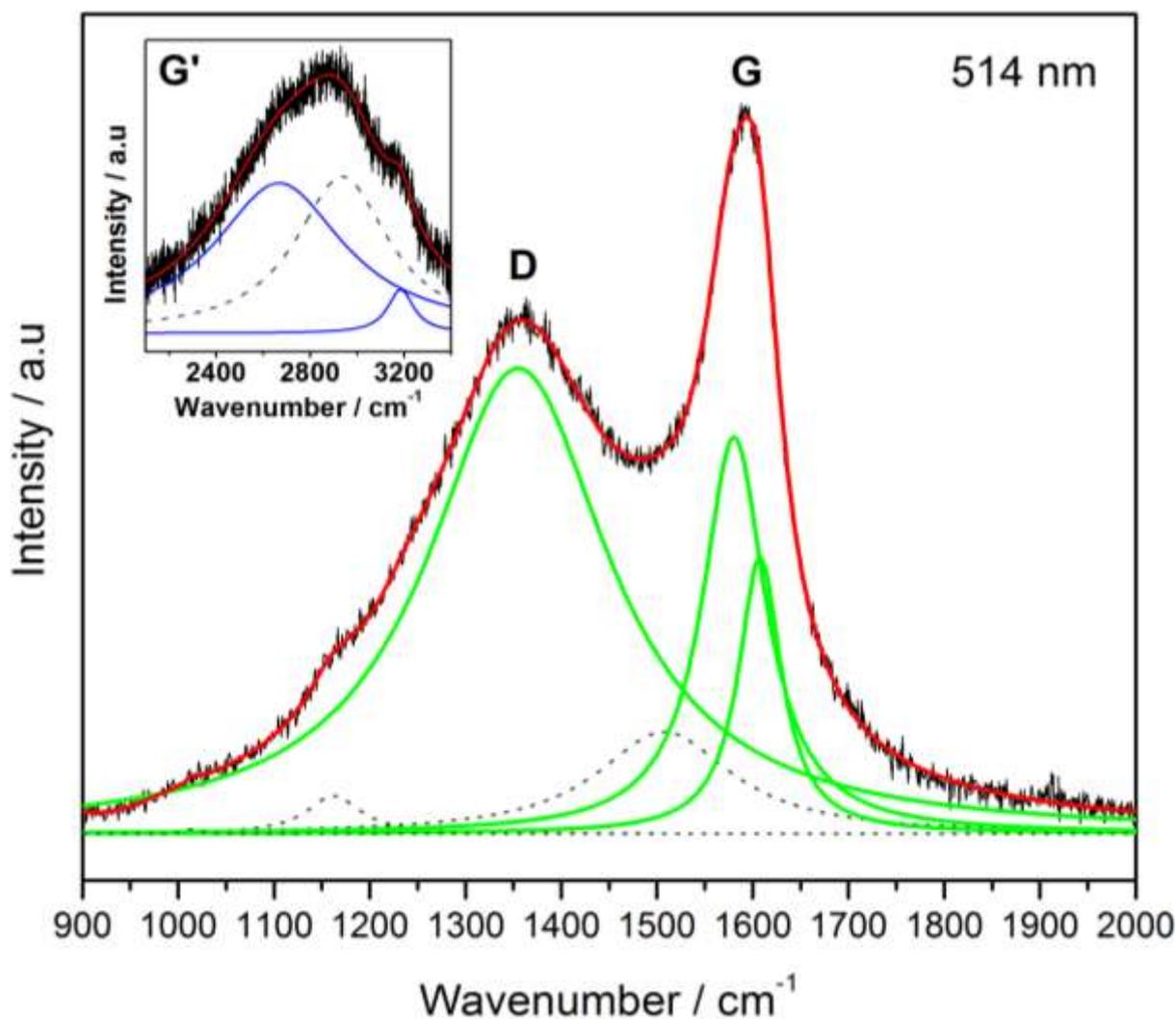

**Figure S9. Raman spectra of CNSs sample.** Lorentzian line-shape peak fitting shown, with dashed lines representing band contributions from ethanol used to disperse the sample. Red outline is the envelope of the peak fitting. Second order peaks, G', shown in the inset.

The Raman spectra of the carbon material best represents a disordered carbon material between nano-crystalline graphite to low $sp^2$ non-crystalline carbon, Figure S9[10], which agreed well with XPS analysis on the ratio of $sp^2$ to $sp^3$ carbon (Table S1). Lorentzian curve fitting employed on the Raman spectrum of the graphene material showed an asymmetry in the 'G band' yielding peaks centered at 1580 cm$^{-1}$ with FWHM 80 cm$^{-1}$ and 1607 cm$^{-1}$ with FHWM 54 cm$^{-1}$ with decreasing relative intensity. This G 'band' is believed to be due to the in-plane stretching motion between pairs of $sp^2$ carbon atoms. This mode does not require the presence of six-fold rings, so it occurs at all $sp^2$ sites not only those in rings, and appears in the range 1500–1630 cm$^{-1}$. The asymmetry in the peak may be caused by doping of the graphitic layers by ethanol present which was used to disperse the sample prior to measurement[11].

The presence of the 'D band' centered at 1355 cm$^{-1}$ with FWHM 230 cm$^{-1}$, is believed to be related to the number of ordered aromatic rings, and affected by the probability of finding a six-fold ring in a cluster. The second order peaks (G') are not well defined, but appear as a small modulated bump between 2200 and 3500 cm$^{-1}$ and best represent a multi-layer graphitic material in the presence of some ethanol.

The intensity ratio of the D band to the G band(s) value, commonly reported as $I_D/I_G$, was 1.2 and 1.7, indicating a significant number of defect sites present.[10,12] This value compares well with reported $I_D/I_G$ values for carbon nanospheres, which range between 0.8–1.2.[13] The relative intensity and positions of the G and D bands have been interpreted to be due to the presence of defects and disorder in the short range graphitic fragments.[14] This is directly verified with TEM (Figure 3 and S3). The identification of bands associated with other phases, which may also be present in smaller quantities (*e.g.* diamond), was not possible due to the background of the Raman spectrum contributions of disordered carbon.

## Supplementary Information – Electron Spin Resonance

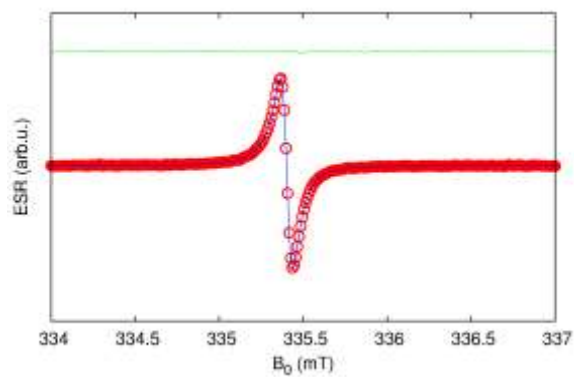

**Figure S10.** Room temperature (300 K) ESR signal from the CNSs testifying the long conduction $T_1=T_2$. A fit to a derivative Lorentzian line-shape (blue line) and the near-zero residual signal (green line) indicating an excellent homogeneous line shape characteristic to itinerant electrons.

For conduction electron spin based qubits the size distribution on the ESR relaxation rate has little effect. This is in contrasting difference from localized paramagnetic spin based q-bits like N@$C_{60}$ N-V centers or other molecular magnet based systems. The ESR relaxation at $E_Z=0$ is limited by $T_1$ because the motional narrowing of conduction electrons is complete. Thus the ESR line is homogeneous and independent of the size distribution. Inhomogeneous broadening comes about at high magnetic fields. The complete motional narrowing of conduction electrons breaks down as electrons progressively confine to cyclotron orbits. This gives the linear broadening by field described by Equation 1. The finite size distribution of the particles induces additional inhomogeneity because the slope of the field dependent broadening depends on the particle size (Equation 2). At the typical ESR frequency at X-band (Figure S11a) the size distribution induced broadening is negligible. At high frequencies the broadening induced by size distribution is enhanced, however, it is still negligible compared to the magnetic field induced broadening as shown in Figure S11b.

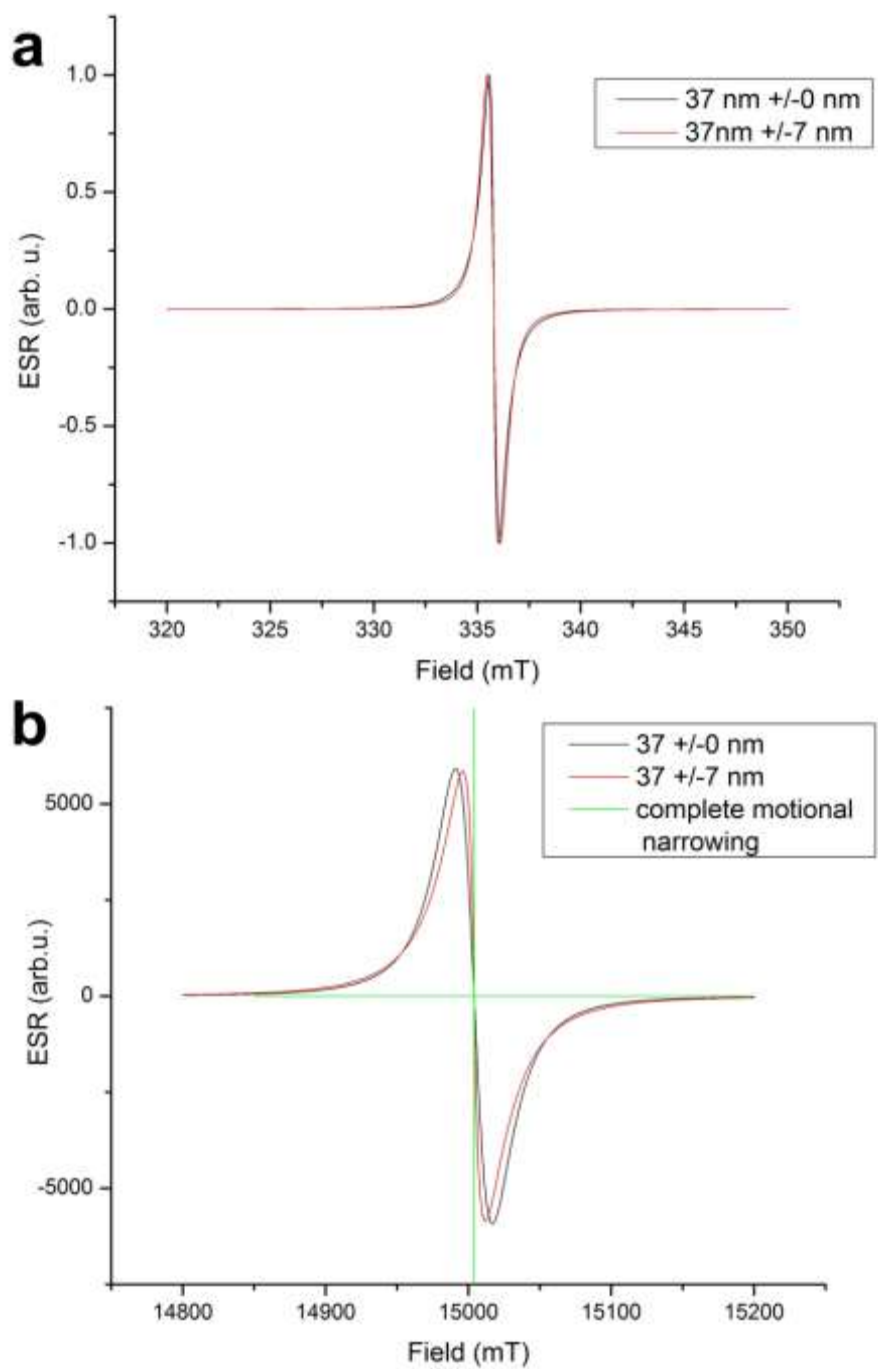

**Figure S11. (a)** Simulated 9.4 GHz ESR spectra of carbon nanospheres based on eq. 1 and 2 with and without the contribution of size distribution, black and red lines respectively. The two spectra are practically identical due to the narrow size distribution. **(b)** Simulated 420 GHz ESR spectra of CNSs based on eq. 1 and 2 with and without the contribution of size distribution, black and red lines respectively. Green line is a simulated spectra assuming (unphysical) complete motional narrowing.

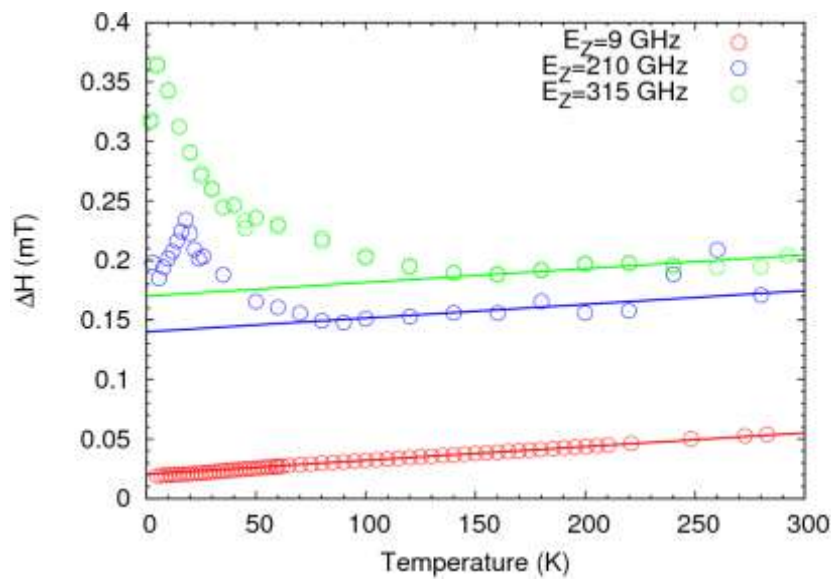

**Figure S12. Temperature dependence of the ESR linewidth measured at different strength of the Zeeman energy.** At high-temperatures where no saturation effect occurs the linewidth decreases linearly by decreasing temperature. This behaviour is commonly observed in metals and explained by spin-orbit coupling by Elliott[15]. The slope is independent of $E_Z$ at high temperatures also in agreement with Elliott mechanism.[15] The deviation from the linear dependence at high-frequencies and at low temperatures is due to ESR saturation effects which occurs progressively at higher temperatures by increasing $E_Z$.[16]